# Crystalline Assemblies and Densest Packings of a Family of Truncated Tetrahedra and the Role of Directional Entropic Forces

Pablo F. Damasceno[1*], Michael Engel[2*], Sharon C. Glotzer[1,2,3†]

[1]Applied Physics Program, [2]Department of Chemical Engineering, and [3]Department of Materials Science and Engineering, University of Michigan, Ann Arbor, Michigan 48109, USA.

[*] These authors contributed equally.

[†] Corresponding author: sglotzer@umich.edu

**ABSTRACT** Polyhedra and their arrangements have intrigued humankind since the ancient Greeks and are today important motifs in condensed matter, with application to many classes of liquids and solids. Yet, little is known about the thermodynamically stable phases of polyhedrally-shaped building blocks, such as faceted nanoparticles and colloids. Although hard particles are known to organize due to entropy alone, and some unusual phases are reported in the literature, the role of entropic forces in connection with polyhedral shape is not well understood. Here, we study thermodynamic self-assembly of a family of truncated tetrahedra and report several atomic crystal isostructures, including diamond, β-tin, and high-pressure lithium, as the polyhedron shape varies from tetrahedral to octahedral. We compare our findings with the densest packings of the truncated tetrahedron family obtained by numerical compression and report a new space filling polyhedron, which has been overlooked in previous searches. Interestingly, the self-assembled structures differ from the densest packings. We show that the self-assembled crystal structures can be understood as a tendency for polyhedra to maximize face-to-face alignment, which can be generalized as directional entropic forces.

Entropic forces are effective forces that result from a system's statistical tendency to increase its entropy. They differ from traditional, conservative ones like van der Waals or Coulomb forces that arise from underlying microscopic interactions. Polymer "elasticity",[1] hydrophobicity[2] and depletion interactions[3] are all examples of entropic forces. Entropy can also cause hard particles, which have no interactions other than their inability to occupy the same region in space, to form crystals and liquid crystals. Hard rods and disks spontaneously align and can order into layers and columns at intermediate packing densities, if those structures increase the configurational space available to the particles.[4,5] Hard spheres crystallize into a face-centered cubic (fcc) structure for the same reason.[6] Recently, the first hard-particle quasicrystal – a complex structure with long-range order but no periodicity, mostly observed in atomic alloys – was discovered in computer simulations of regular tetrahedra.[7] Beyond simple crystals and liquid crystalline phases, experimental realizations of complex colloidal crystals, such as those isostructural to $AB_n$ atomic crystals, have required two-component mixtures or attractive interactions,[8-10] regardless of building block shape. Finding new and alternative ways to assemble novel superstructures is an important prerequisite for future applications of nanoparticles and colloids.

Here we investigate the influence of shape on particle self-assembly and show that small changes in shape of hard polyhedra can suffice to produce a range of colloidal crystal structures whose rich complexity rivals that of atomic analogues, without the need for mixtures or attractive interactions. To demonstrate this, we study *via* computer simulation assemblies and packings of hard tetrahedra as their corners are increasingly and symmetrically truncated until the particles become octahedra. In addition to body-centered cubic and quasicrystalline structures, we report a space-filling polyhedron forming β-tin, a densest packing for the Archimedean truncated tetrahedron into α-arsenic, and the self-assembly of crystals isostructural to diamond and high-pressure lithium. We show that in all cases the assembled structures favor face-to-face alignment, and, building on earlier concepts of entropy-driven phase transitions,[4,5] we introduce the notion of directional entropic forces for hard particles in analogy with controlled entropic forces in systems of colloids and with directional enthalpic forces in molecular and patchy particle systems. This notion explains the observed crystal superlattices in a natural way.

Our choice of building blocks is motivated by recent advances in the synthesis and assembly of faceted and monodisperse nanocrystals,[11-17] and especially tetrahedra.[18,19] The observation that the structure of many atomic crystals is dominated by polyhedral building blocks[20] suggests the possibility to reproduce atomic crystal isostructures with colloidal polyhedra. However, the phase behavior of polyhedra is often unknown even for the highly symmetric Platonic and Archimedean solids. Exceptions are mesophases[21] and phases in the limit of high pressure, which are identical to maximally dense packings and have been studied as a mathematical optimization problem.[22-26] To date, no large-scale assembly simulations of complex crystal structures with thousands of polyhedra have been reported in the literature.

We use Monte Carlo computer simulation as in Ref. 7. To identify a crystal structure, each polyhedron is replaced by an atom positioned at its centroid. We consider this replacement when we say that an arrangement of polyhedra is "isostructural" to an atomic crystal. The amount of truncation applied to a tetrahedron is specified by the truncation parameter $t$. This parameter increases linearly with the distance of a truncation plane from the nearest of the four vertices in the original regular tetrahedron. A truncated tetrahedron has four equilateral triangles with edge length $\sigma(t/2)$ and four hexagons with the two edge lengths $\sigma(1-t)$ and $\sigma(t/2)$. Special cases depicted in Fig. 1a are the regular tetrahedron, with $t = 0$, the (as we will see) space-filling truncated tetrahedron (STT), with $t = 1/2$, the Archimedean truncated tetrahedron (ATT), with $t = 2/3$, and the regular octahedron, with $t = 1$.

## Results and Discussion.

**Densest packings.** We first investigate the densest packings as a function of truncation. The findings are later compared to the self-assembled structures. Results for the maximum packing density $\phi_{max}$ are included in Figs. 1b and 2a. We find that $\phi_{max}(t)$ is continuous, but shows sharp kinks.[26] The observation that all truncated tetrahedra pack with densities of at least $\phi_{max}$ = 82% indicates that these shapes are generally efficient packers. They pack much better than spheres, which have a maximum packing density of only 74%. The number of particles per primitive unit cell jumps from $n$ = 4 for weakly truncated tetrahedra ($t$ < 0.27), to $n$ = 2 for intermediate cases (0.27 < $t$ < 0.93), and finally to $n$ = 1 for strongly truncated tetrahedra resembling octahedra ($t$ > 0.93). This decrease in $n$ is expected because shapes tend to pack into simpler configurations with fewer particles per unit cell if they are more centrally symmetric.[24] We observe that neighboring particles always pack face-to-face with at least partial registry. For $n$ = 2, the orientations of the two truncated tetrahedra are related by inversion. For $n$ = 4, the particles form two dimers which are again related by inversion.

A further classification is possible by studying the geometry of the primitive unit cell with varying $t$. We observe that the unit cell of the densest packing shears smoothly with $t$ when contact points between neighbors can be maintained and slide along the particle surfaces. A kink in $\phi_{max}(t)$ appears if there is a denser packing with a different neighbor configuration, or if a contact point hits an edge of a polyhedron as $t$ changes, in which case deformations require the unit cell to shear in another direction. The lattice parameters of the densest packings are shown in Fig. 2b. The variation of the lattice parameters with truncation exhibits several transitions. Symmetric packings (monoclinic, tetragonal, orthorhombic) occur for intermediate truncations. Shapes close to the octahedron favor triclinic lattices. As reported previously, perfect tetrahedra arrange as dimers and pack most densely in a triclinic double dimer packing[25,26] and octahedra pack most densely in a rhombohedral Bravais lattice.[22]

New packings appear for $t$ = 1/2 and $t$ = 2/3, both local maxima of $\phi_{max}(t)$ in Fig. 1b. The STT at $t$ = 1/2 (Fig. 2d) has $\phi_{max}(t)$ = 1 and is therefore space filling. The centroids of the STT form the $\beta$-tin structure (Supplementary Information, Fig. S1). The exact analytical positions and orientations of the STT in this structure are given in Table S1. It appears the STT tiling has been missed in previous searches[27] and thus the STT is a previously unrecognized space-filling polyhedron. Given the simplicity and high symmetry of the STT and the sparsity of space-filling polyhedra in general, this discovery is unexpected.

The importance of the packing at $t = 2/3$ lies in the regularity of the ATT. As an Archimedean solid, the ATT has four regular hexagonal faces and four regular triangular faces. Based on the findings in our simulations, we analytically construct an ideal tiling of ATT with packing density $\phi_{max} = 207/208$ (Fig. 2e). A similar packing density has also recently been reported[28] in parallel to this work. Other previous studies have reported packings with densities $\phi_{max} = 23/24$ [24] and $\phi_{max} = 0.988...$[29] for the ATT. Although the number of the densest ATT packing is surprisingly close to 1, the ATT does not fill space because of the appearance of small tetrahedral voids with edge length $\sigma/9$. The centroids of the ATT form the $\alpha$-arsenic structure (Fig. S2). The exact analytic construction of the ATT packing is given in Table S2.

**Self-assembly of crystal structures.** Next we study the self-assembly of hard truncated tetrahedra into crystals starting from the disordered fluid phase. We choose a truncation $t$ and an initial packing density $\phi$ for a constant volume Monte Carlo simulation and simulate for 20-100 million of Monte Carlo cycles. The lowest packing density where we observe crystallization, $\phi_c$, is shown as the lower curve in Fig. 1b. Crystallization does not occur on the timescale of our simulations in the region $0.8 < t < 0.85$, indicating that either the nucleation rate is extremely low, or crystals are not possible. For all other ranges of $t$ between 0 and 1, we repeatedly observe nucleation and rapid growth of single-crystal phases spanning the whole simulation box. We find that $\phi_c$ varies between 0.5 and 0.63 depending on shape and increases towards the boundaries of the stability regimes.

Five crystal structures spontaneously self-assemble in our simulations as $t$ is increased from zero. Over a considerable range of truncation ($0 \leq t \leq 0.45$), the particles order into a dodecagonal quasicrystal (Fig. 3a). This shows that the quasicrystal reported previously with regular tetrahedra[7] is robust and forms even with truncated particles. However, as $t$ increases towards 0.5, crystallization slows and the quality of the quasicrystal deteriorates. For $0.5 \leq t \leq 0.8$, the diamond structure assembles from the fluid. Nucleation is fast and crystal growth is robust enough to obtain defect-free single crystals routinely with thousands of particles (Fig. 3b). Towards the higher end of this range of $t$, the cubic unit cell of diamond compresses along one direction to form the tetragonal crystal structure β-tin, identical to the densest packing of the STT, but for different values of $t$.

Strongly truncated tetrahedra with $0.85 \leq t < 0.95$ assemble into a cubic phase isostructural to high-pressure lithium.[30] The lattice is bipartite and comprised of two identical but intertwined sublattices, as indicated by the coloring in Fig. 3c. Although the structure is complex with 16 particles in the unit cell, perfect single crystals assemble easily from the fluid (Fig. S5). To our knowledge, crystals with such a high number of particles in the unit cell have so far not been reported for hard particles in the literature. For nearly perfect octahedra ($0.95 \leq t \leq 1$), the body-centered cubic (bcc) lattice is observed at low density (Fig. 3d). This is expected because nano-octahedra with short-range repulsive interactions self-assemble into bcc.[31]

It is interesting to note that the densest packing is different from the stable phase nucleating from the fluid (Fig. 1b). We do not find even one value for the truncation parameter $t$, where the densest packings and assembled crystals are identical. Even in the case of perfect octahedra, the bcc crystal at low density transforms into a rhombohedral lattice by a slight shear along the [111] direction. This indicates that the self-assembly of hard truncated tetrahedra, and possibly of a large class of polyhedra, cannot be viewed as a packing problem alone. We further observe that the self-assembled structures typically have higher crystallographic symmetry (tetragonal, cubic, dodecahedral) compared to their corresponding densest packings (triclinic, monoclinic, tetragonal, orthorhombic) and have more face-to-face orientations between nearest neighbors.

**Directional entropy.** Indeed, face-to-face arrangements appear to be a key motif in the self-assembly of hard polyhedra. The crystals we observe can be grouped into three classes with varying symmetries and varying importance of face-to-face orientations: (i) bcc-like structures, (ii) a tetrahedral network, and (iii) carbon-related structures. The bcc-like structures (bcc, high-pressure lithium) result for particles with shapes resembling the octahedron. The high-pressure lithium phase relates closely to bcc by shifting atom columns.[30] Within each sublattice of the bipartite lithium structure there are hexagon-hexagon face-to-face contacts only (Fig. 3c), while triangle-hexagon and triangle-triangle orientations occur only between sublattices (Fig. S4). It is interesting that the high-pressure lithium phase has recently also been observed with attractive octahedral nanocrystals,[32] which demonstrates that in this example changing the shape of the octahedron to a truncated tetrahedron has a similar effect on the preferred crystal structure as turning on an attractive face-to-face interaction. In bcc, the higher equivalence of triangle and hexagon faces facilitates two types of face-to-face orientations (Fig. S6), explaining the transition from high-pressure Lithium to bcc for increasing values of the truncation parameter.

At the other extreme, the dodecagonal quasicrystal can be understood as a tetrahedral network of pentagonal dipyramids.[7] It occurs for particles with shape most closely resembling the tetrahedron and is dominated by triangular face-to-face contacts that produce dimers arranged in stacks of pentagonal dipyramids and 12-member rings. Shapes intermediate between the tetrahedron and the octahedron form a range of carbon-like structures, including diamond, β-tin, and α-arsenic, with strong face-to-face alignment of hexagonal facets. In contrast to the tetrahedral network, neighboring faces are now rotated by 180 degrees relative to each other, which permits a higher contact area for these truncations (Fig. S3).

Thermodynamic systems minimize free energy at equilibrium; when interactions are absent, entropy is maximized at fixed volume. The favoring of face-to-face orientations provides insight into how entropy is maximized by the structures we observe. Consider the well-known, isotropic-to-nematic transition in hard rods.[4] At sufficiently high packing densities, parallel alignment increases the translational entropy of the system at the expense of rotational entropy, producing the nematic phase.[5] Simple analogy suggests that polyhedra should likewise prefer to align with faces parallel, in agreement with our observations. Competition among different types of faces, as in the present family of polyhedra, can lead to complexity of structure.

## Conclusion.

Our simulation results constitute the first report of the self-assembly of complex crystal structures from the fluid phase with hard particles. The findings can be explained by a preference for face-to-face alignment in systems of hard, faceted objects, which suggests considering entropic forces as directional in such systems, since they increase with contact area between neighbors[3] and thus act strongest perpendicular to the faces. In this sense, directional entropic forces in hard particle systems act like attractive interactions in systems of patchy particles, or like depletion interactions in mixtures of large and small particles, which can be controlled by shape [17,33] and roughness.[34] Our results suggest that particle shape can produce entropic "patchiness" just as patterning can produce interaction patchiness[35] and that bond geometries in patchy particle systems can, in principle, be imitated by shape, providing additional strategies for building block design. A recent example of directional entropic forces in colloids are dimpled "pacman" particles,[33] in which the dimple shape and entropic depletion forces control particle binding. The use of depletants with polyhedral particles should provide an even greater variety of ordered structures, obtained then by entropic patchiness arising both from facet alignment and anisotropic depletion.

## Methods.

**Simulations.** Our numerical tools are isochoric (constant volume) and isobaric (constant pressure) Monte Carlo simulations similar to the ones employed in Ref. 6. Update steps consist of arbitrary translation, rotation and box shearing moves. Step sizes are chosen to achieve an average acceptance probability of 30%. While the simulation box is in principle allowed to have arbitrary dimensions and to undergo shear in any of three orthogonal directions, we use a lattice reduction technique after each box move to keep the box as compact as possible and avoid strong shearing. Candidates for densest packings are obtained by compressing systems with 1, 2, or 4 particles in the box applying periodic boundary conditions. Self-assembly simulations employ much larger boxes with at least 1000 particles and up to 8000 particles at packing densities $0.48 \leq \phi \leq 0.66$. All findings are verified by running independent simulations with different initializations and in different box sizes.

**Structure characterization.** For the determination of radial distribution functions and the calculation of diffraction patterns *via* fast Fourier transform, point scatterers are placed in the centroids of the polyhedra. Bond order diagrams are obtained by projecting the vectors connecting the centroids of nearest neighbor polyhedra on the surface of a sphere. All three functions are used to characterize and identify the self-assembled structures and densest packings.

**Geometric representation.** The position and orientation of a truncated tetrahedron are given by a translation vector ($x$, $y$, $z$) and a quaternion ($a$, $b$, $c$, $d$). The relation between the quaternion and the rotation matrix is:

$$R = \begin{pmatrix} a^2 + b^2 - c^2 - d^2 & 2(bc - ad) & 2(bd + ac) \\ 2(bc + ad) & a^2 - b^2 + c^2 - d^2 & 2(cd - ab) \\ 2(bd - ac) & 2(cd + ab) & a^2 - b^2 - c^2 + d^2 \end{pmatrix}.$$

**Acknowledgements.**
This work was supported in part by the U. S. Air Force Office of Scientific Research (FA9550-06-1-0337) and by a U.S. Department of Defense National Security Science and Engineering Faculty Fellowship (N00244-09-1-0062). M.E. acknowledges support from the Deutsche Forschungsgemeinschaft (EN 905-1/1).

**Supporting Information Available.**
Supporting Information Available: Details for the analytic construction of the densest packings and self-assembled crystal structures. This material is available free of charge *via* the Internet at http://pubs.acs.org.

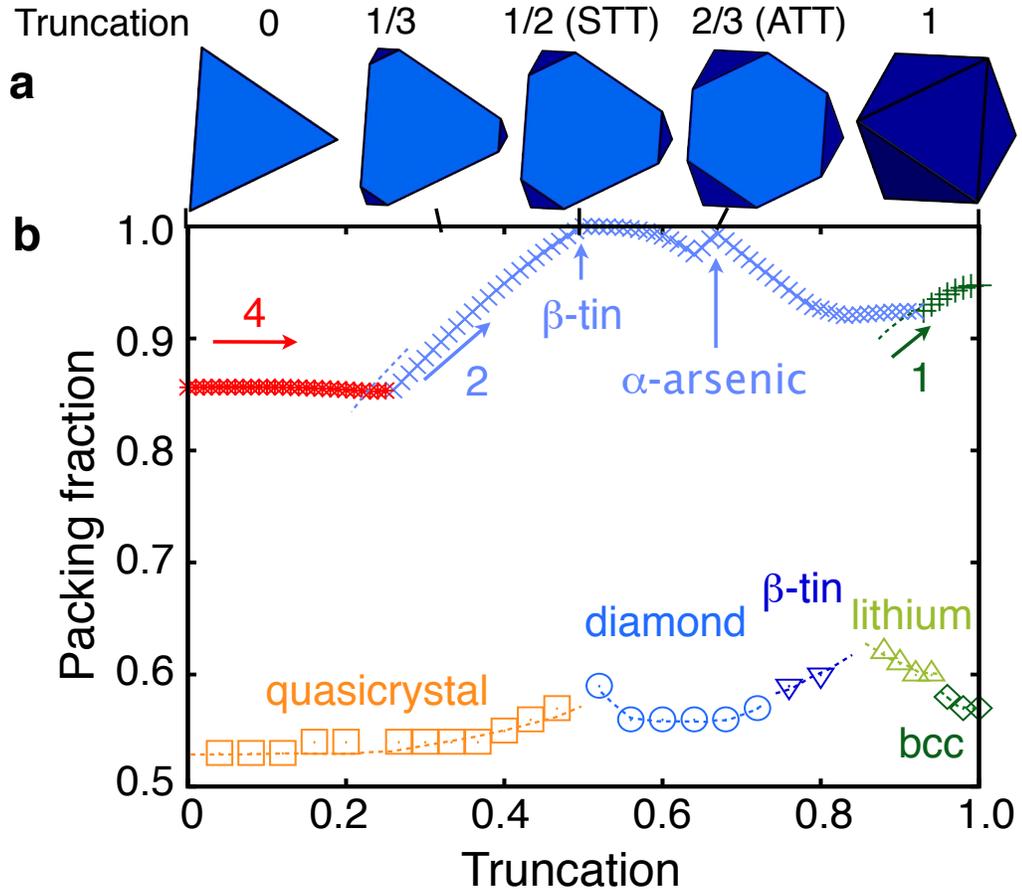

**Figure 1. Dense packings and self-assembled truncated tetrahedra into atomic crystal isostructures. a,** The family of truncated tetrahedra is parameterized by the truncation parameter $t$ and ranges from the regular tetrahedron to the regular octahedron. **b,** Phase diagram of truncated tetrahedra. Upper datasets indicate maximum packing densities $\phi_{max}$. Numerals and corresponding colors indicate the number of particles in the primitive unit cell. Labels indicate β-tin and α-arsenic structures for the space-filling and Archimedean truncated tetrahedra, respectively. Lower datasets corresponds to the lowest density, $\phi_c$, at which crystallization from the fluid is observed. Labels indicate the dodecagonal quasicrystal, diamond, β-tin, high-pressure lithium, and body-centered cubic structures.

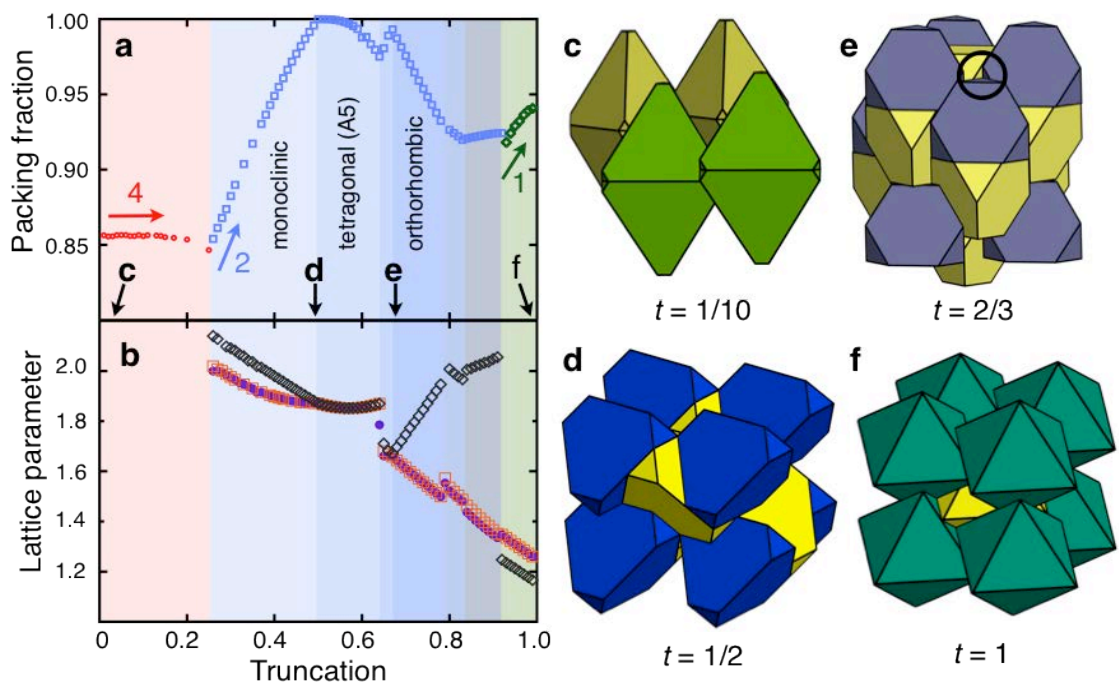

**Figure 2. Densest packings of truncated tetrahedra. a,** Blow-up of maximum packing densities from Fig. 1b and comparison with **b,** lattice parameters vs. truncation $t$. Colors of the data sets in (a) (red, blue green) indicate the number of particles in the primitive unit cell (four, two, one) for indicated structures. Colors of the data sets in (b) (magenta, dark blue, dark green) indicate the absolute value of the three lattice parameters that generate the unit cell. In both (a) and (b), vertical background shading indicates unit cells with different numbers of particles or different symmetries (vertical regions inside blue shading). Discontinuities of the derivative of $\phi_{max}$ occur at $t = 0.50, 0.64, 2/3, 0.78, 0.83$ and $0.94$, and indicate changes in symmetry of the densest packing structure. These kinks are reflected by similar discontinuities in the lattice parameter curve (or its derivative) in (b). **c,** For small truncations the particles form dimers, which occur in two orientations similar to the densest tetrahedron packing. **d,** The STT fills space in the β-tin structure. Two triangular faces fit along a long edge of the hexagonal faces. **e,** The ATT packs with a density of 207/208 in a lattice isostructural to α-arsenic. The particles are perfectly face-to-face within dimers, but not with other neighbors. **f,** Regular octahedra pack into a rhombohedral Bravais lattice, which is slightly sheared compared to the bcc lattice.

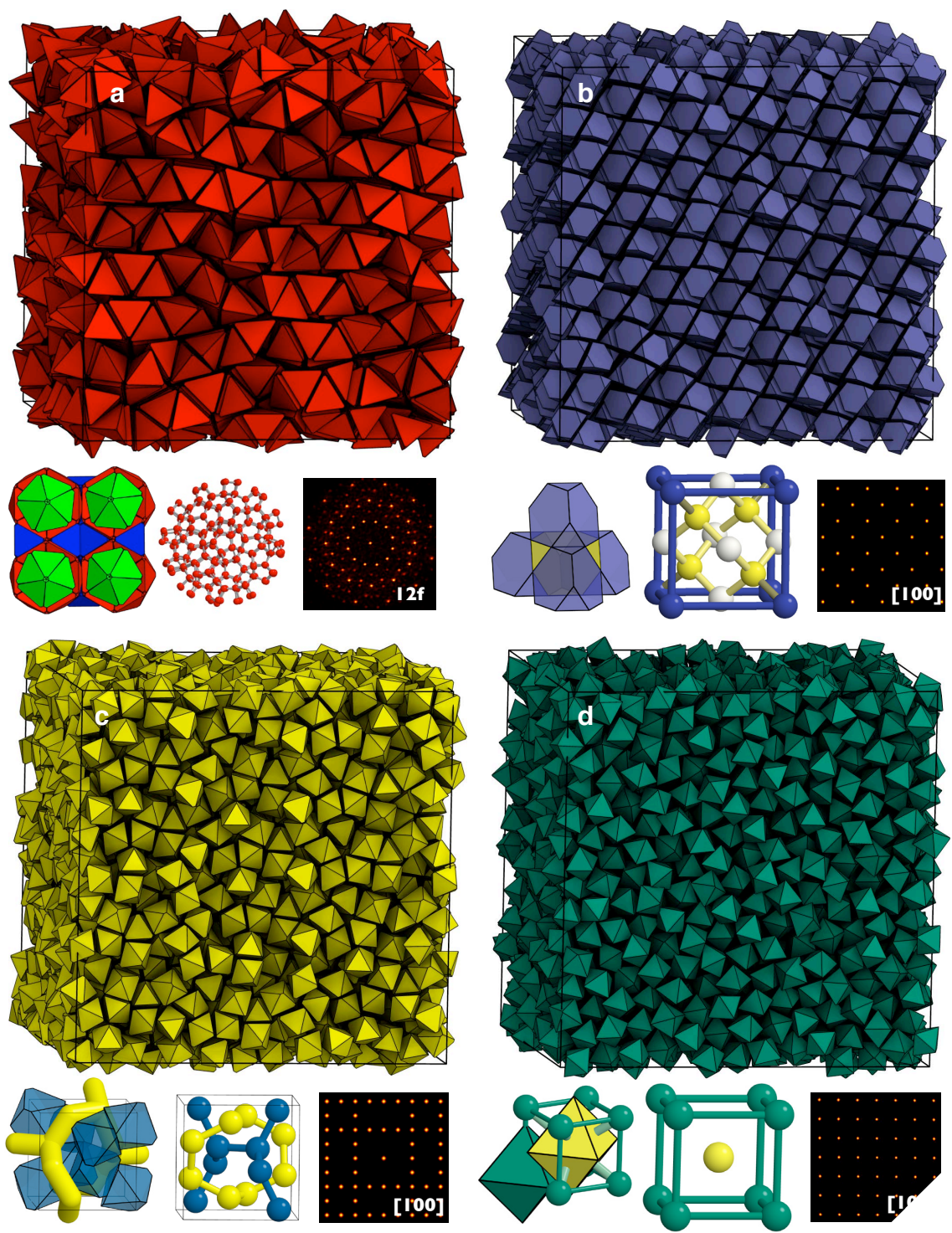

**Figure 3. Structures of truncated tetrahedra self-assembled in simulation at intermediate density**. In each subfigure a snapshot of the full simulation box, temporally averaged to remove thermal disorder together with a characteristic motif (bottom left),

ball-and-stick model (bottom center) and the system's diffraction pattern (bottom right) are shown. With increasing truncation, 2624 truncated tetrahedra assemble into **a,** dodecagonal quasicrystal ($t = 0.2$), **b,** diamond lattice ($t = 2/3$), **c**, bipartite lattice isostructural to high-pressure lithium ($t = 0.95$), and **d,** bcc lattice of regular octahedra ($t = 1$). Not shown: β-tin ($t = 0.8$), which is the same structure as that obtained for the densest STT structure at a different value of $t$.

# Crystalline Assemblies and Densest Packings of a Family of Truncated Tetrahedra and The Role of Directional Entropic Forces

## – Supplementary Information –

*Pablo F. Damasceno[1*], Michael Engel[2*], Sharon C. Glotzer[1,2,3†]*

[1] Applied Physics Program, [2] Department of Chemical Engineering, and [3] Department of Materials Science and Engineering, University of Michigan, Ann Arbor, Michigan 48109

[*] These authors contributed equally.
[†] Corresponding author: sglotzer@umich.edu

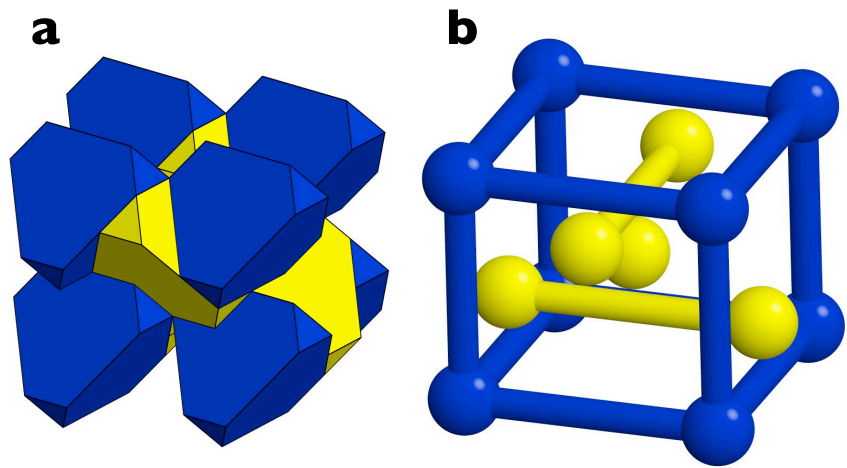

**Figure S1 | The packing of space filling truncated tetrahedra (STT). a,** Snapshot of the particles arranged into the β-tin structure (Strukturbericht designation A5, tetragonal diamond) with lattice parameters of the non-primitive cell $a = 1.6^{-1/2}s$ and $c = 2.0^{-1/2}s$, and space group I4$_1$/amd (No. 141). Note that the edge lengths of the hexagonal facets have a ratio of 2:1, which allows neighboring STT to align effectively along edges. The STT tiling is intricate because neighboring polyhedra are only in partial face-to-face contact. **b,** Ball-and-stick model of β-tin.

| a | Lattice Vectors | $b_1 = (6, -2, -4)$; $b_2 = (-2, -6, 4)$; $b_3 = (-6, 2, -4)$ |
|---|---|---|
| b | Vertices of the space-filling truncated tetrahedron | $v_1 = (4, 2, 2)$; $v_2 = (2, 4, 2)$; $v_3 = (2, 2, 4)$; $v_4 = (4, -2, -2)$; $v_5 = (2, -2, -4)$; $v_6 = (2, -4, -2)$; $v_7 = (-2, 4, -2)$; $v_8 = (-2, 2, -4)$; $v_9 = (-4, 2, -2)$; $v_{10} = (-2, -2, 4)$; $v_{11} = (-2, -4, 2)$; $v_{12} = (-4, -2, 2)$ |
| c | Particle 1 | position = (2, 1, 1); orientation = (1, 0, 0, 0) |
| d | Particle 2 | position = (-2, -1, -1); orientation = (cos(p / 4), sin(p / 4), 0, 0) |

**Table S1 | Structural data of the packing of space filling truncated tetrahedra (STT) with two particles in the unit cell. a,** Lattice vectors spanning the unit cell. **b,** The twelve vertices of the STT in standard orientation positioned at the origin. **c,d,** Positions and orientations of the two particles.

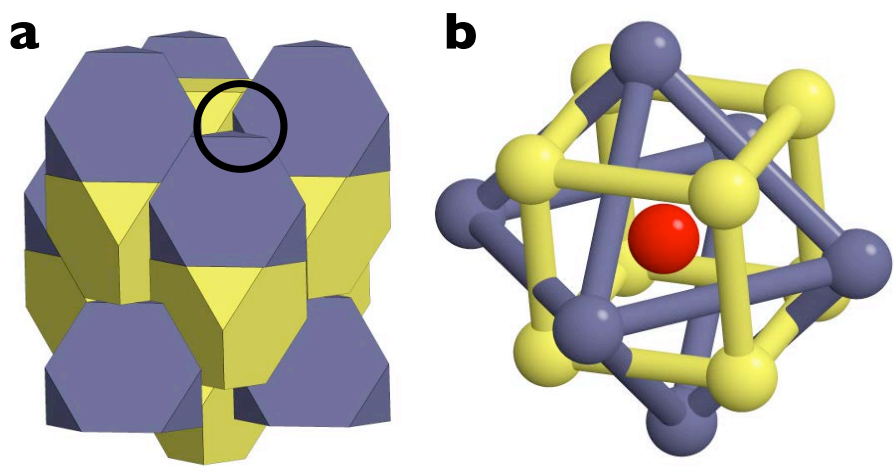

**Figure S2 | The densest packing of the Archimedean truncated tetrahedron (ATT) with packing density $f_{max} = 207/208$. a,** Snapshot of the particles arranged into the α-arsenic structure (Strukturbericht designation A7) with three lattice parameters $a \approx 0.3984s$, $b \approx -0.1682s$, $u = 72^{-1/2}s$ and space group $R\bar{3}m$ (No. 166). The packing is not space-filling because of small tetrahedral voids found between the dimers. One such void is indicated by a black circle. **b,** Ball-and-stick model of α-arsenic.

| a | Lattice Vectors | $b_1 = (-4, 8, 12)$; $b_2 = (12, -4, 8)$; $b_3 = (8, 12, -4)$ |
|---|---|---|
| b | Vertices of the Archimedean truncated tetrahedron | $v_1 = (9, 3, 3)$; $v_2 = (3, 9, 3)$; $v_3 = (3, 3, 9)$; $v_4 = (9, -3, -3)$; $v_5 = (3, -3, -9)$; $v_6 = (3, -9, -3)$; $v_7 = (-3, 9, -3)$; $v_8 = (-3, 3, -9)$; $v_9 = (-9, 3, -3)$; $v_{10} = (-3, -3, 9)$; $v_{11} = (-3, -9, 3)$; $v_{12} = (-9, -3, 3)$ |
| c | Particle 1 | position = $(3, 3, 3)$; orientation = $(1, 0, 0, 0)$ |
| d | Particle 2 | position = $(-3, -3, -3)$; orientation = $(\cos(\pi/4), \sin(\pi/4), 0, 0)$ |

**Table S2 | Structural data of the packing of Archimedean truncated tetrahedra (ATT) with two particles in the unit cell. a,** Lattice vectors spanning the unit cell. **b,** The twelve vertices of the ATT in standard orientation positioned at the origin. **c,d,** Positions and orientations of the two particles.

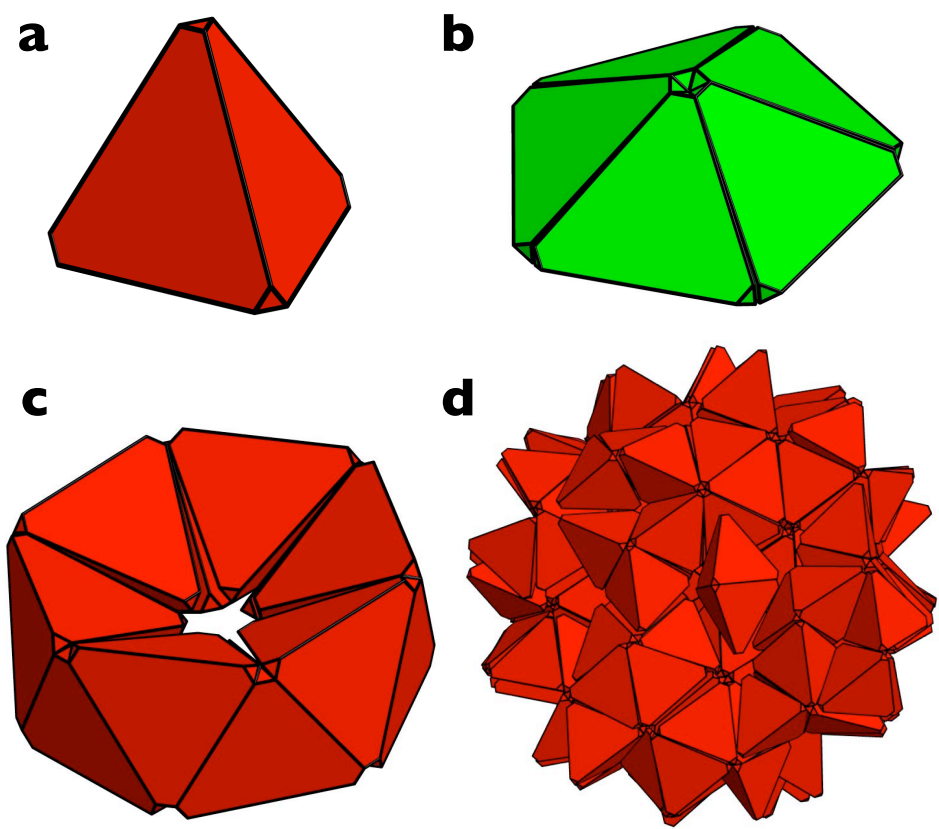

**Figure S3 | Hierarchical assembly of slightly truncated tetrahedra into a dodecagonal quasicrystalline structure. a,** Slightly truncated tetrahedra ($t = 0.1$) form **b,** pentagonal dipyramids of five particles and **c,** rings consisting of twelve particles. These local building blocks ultimately assembly into a 12-fold quasicrystal. A small patch of the quasicrystal is shown in **d**.

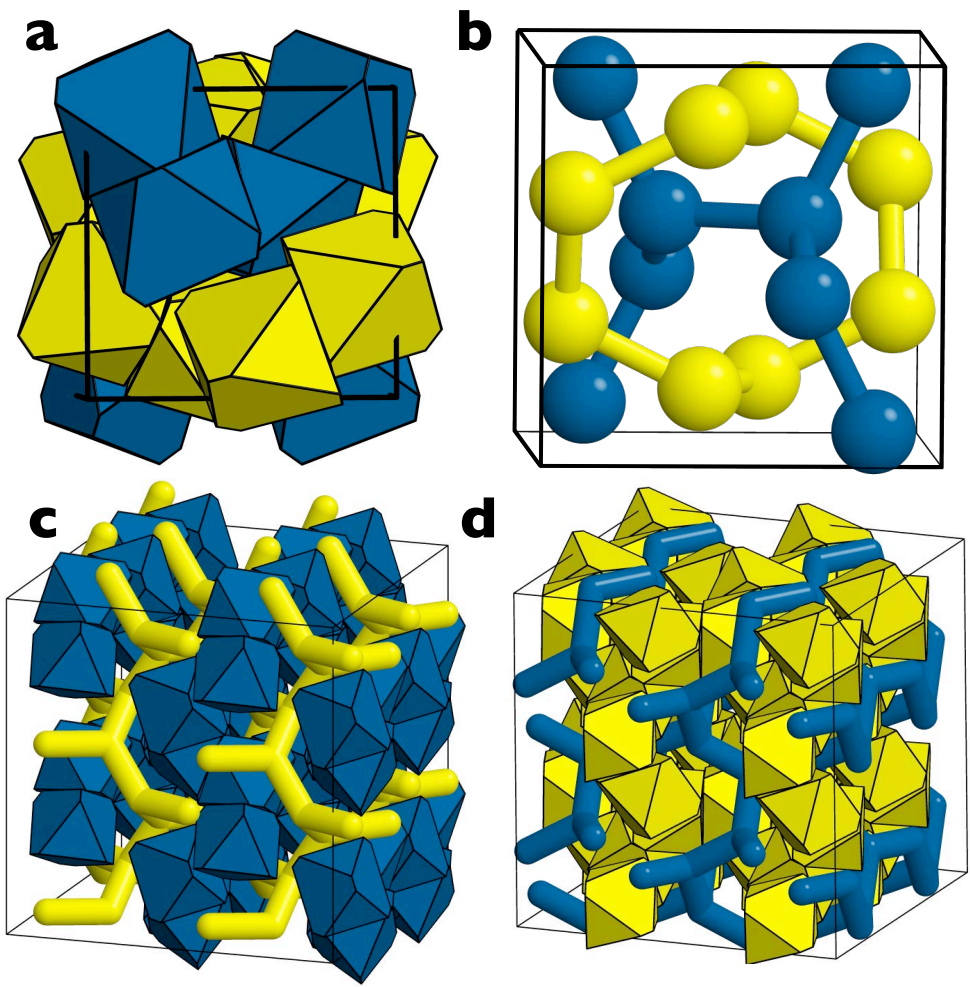

**Figure S4 | Highly truncated tetrahedra ($t = 0.9$) resembling octahedra self-assemble into a phase isostructural to high pressure lithium. a,** Snapshot of the cubic unit cell. Each truncated tetrahedron is connected to three nearest neighbors along its hexagonal faces. The connection splits the structure up into two sublattices as indicated by the coloring (yellow and blue). **b,** The centroids of the nearest neighbors are connected with bonds showing the two sublattices. **c,d,** The two sublattices shown separately. The sublattices are not connected to each other via nearest neighbor bonds.

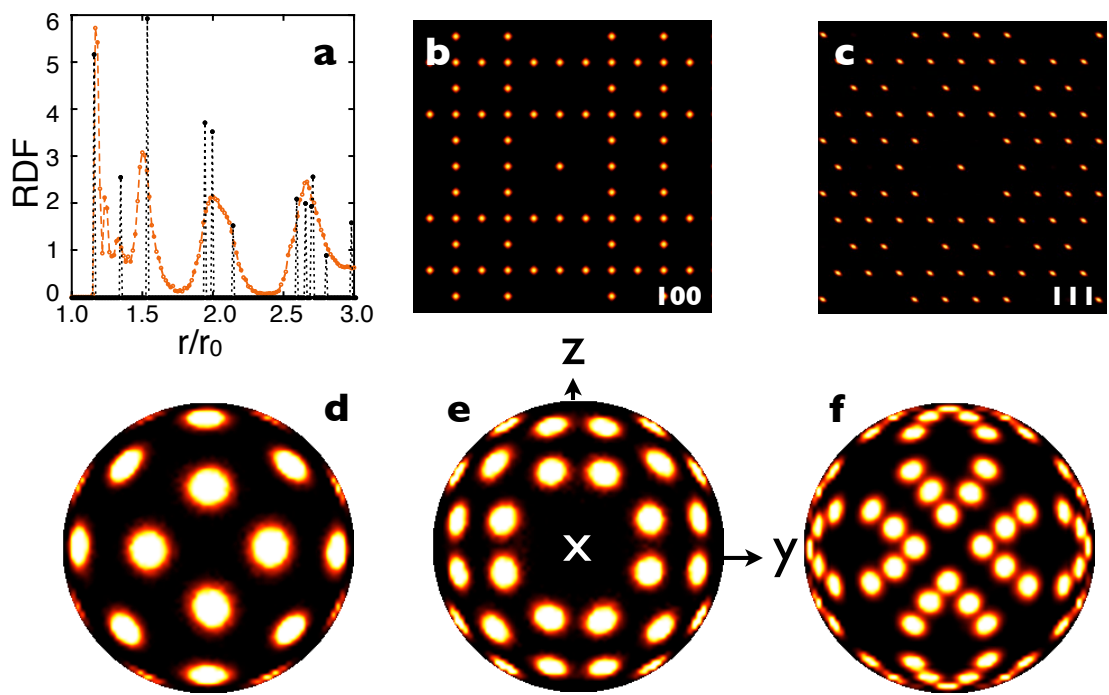

**Figure S5 | Structural characterization of the phase isostructural with high pressure lithium. a,** Radial distribution function of a perfect (black) and self-assembled system (orange) composed of 2624 truncated tetrahedra. **b-c,** Diffraction pattern along [100] and [111] crystallographic directions. **d-f,** Bond order diagrams along the x-direction. The diagrams correspond to the neighbors found within a distance covered by the first peak, $1.0 < r < 1.4$ (d), the second peak, $1.4 < r < 1.75$ (e), and the third peak, $1.75 < r < 2.4$ (f) of the radial distribution function.

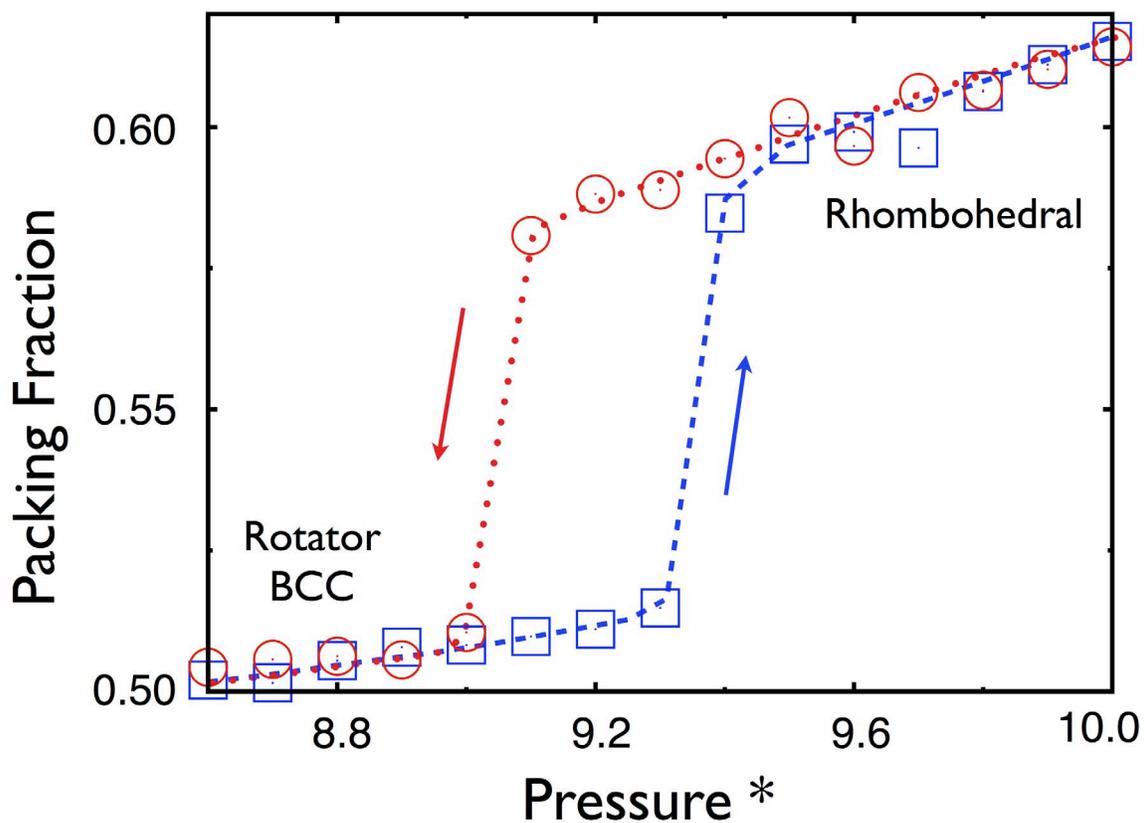

**Figure S6 | Equation of state calculated for perfect octahedra.** Equation of state from simulations of perfect octahedra (t=1). The blue points represent systems of 1024 particles initially in a rotator bcc phase that undergoes a first-order phase transition to the rhombohedral phase under compression. The red line, conversely, represents systems initially in the rhombohedral packing that transform into the rotator bcc phase under decompression.

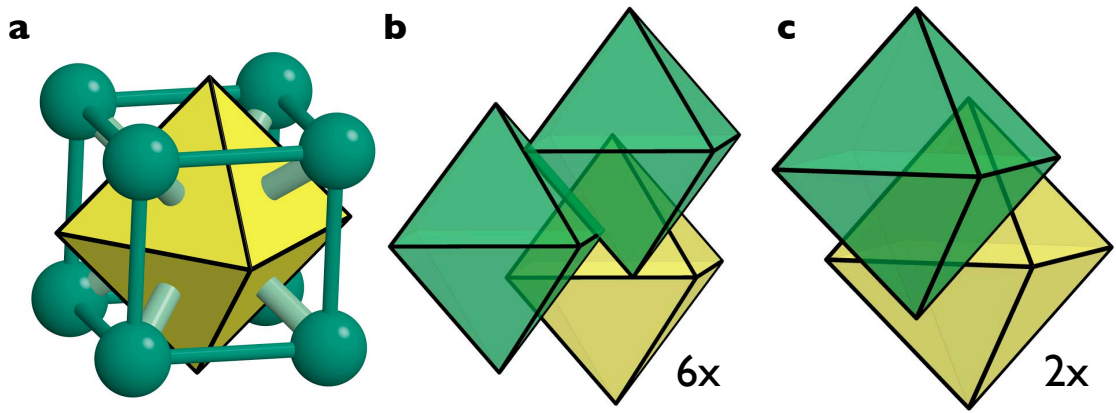

**Figure S7 | Arrangement of octahedra into a weakly sheared bcc lattice. a,** The orientation of the octahedra breaks the cubic symmetry. Six of its eight faces align face-to-face with neighbors as shown in **b**. The two remaining phases align as shown in **c**.